# LIMITS TO THE MASS AND THE RADIUS OF THE COMPACT STAR IN SAX J1808.4–3658 AND THEIR IMPLICATIONS

Sudip Bhattacharyya[1,2]

## ABSTRACT

We show that a survey of equations of state and observations of X–ray pulsations from SAX J1808.4–3658 give 2.27 $M_\odot$ as the upper limit of the compact star mass. The corresponding upper limit of the radius comes out to be 9.73 km. We also do a probabilistic study to estimate the lower limit of the mass of the compact star. Such a limit puts useful constraints on equations of state. We also discuss the implications of the upper mass limit for the the evolutionary history of the source, as well as the detection of it in radio frequencies. We envisage that the possible observation of radio–eclipse may be able to rule out several soft equation of state models, by setting a moderately high value for the lower limit of inclination angle.

*Subject headings:* accretion, accretion disks — binaries: close — equation of state — pulsars: individual (SAX J1808.4–3658) — X-rays: stars

## 1. INTRODUCTION

The discovery of millisecond X–ray pulsations (period $T = 2.49$ ms; Wijnands & van der Klis 1998) in the transient X–ray burster SAX J1808.4–3658 confirmed the speculation that LMXBs are progenitors of millisecond pulsars (Bhattacharya & van den Heuvel 1991). The orbital period ($P_{\rm orb} = 2.01$ hr) and the pulsar mass function ($f_1 = 3.7789 \times 10^{-5}$) of this source were observationally determined by Chakrabarty & Morgan (1998). These give valuable information about the masses (of both the primary and the secondary) and the inclination angle. For example, the value of $P_{\rm orb}$ uniquely determines the mass of a Roche lobe filling low mass star with known mass–radius relation.

It has been recently proposed that the compact star in SAX J1808.4–3658 is a strange star (SS) and not a neutron star (NS) (Li et al. 1999). Such a speculation, if confirmed, will prove that the so called *strange matter hypothesis* (Witten 1984) is correct. According to this hypothesis, strange quark matter (made entirely of deconfined $u$, $d$ and $s$ quark) could be the true ground state of strongly interacting matter rather than $^{56}$Fe. This is an important problem of the fundamental physics. To resolve it, we need to constrain the equations of state (EOS) for this compact star very effectively.

In this Letter, we estimate the upper limits of the mass and the radius of the compact star in SAX J1808.4–3658. We also discuss the possible ways to estimate lower mass limit.

## 2. UPPER LIMITS TO MASS AND RADIUS

We estimate the upper limits of the mass and the radius of the compact star in SAX J1808.4–3658 by using the basic requirements for X–ray pulsations. Here, upto eqn. 5, we follow the same method as described in Li et al. (1999). To explain it, we first define the corotation radius ($R_{\rm co}$) and the magnetospheric radius ($R_{\rm mag}$). They are given by (see Burderi & King 1998)

$$R_{\rm co} = 1.5 \times 10^6 m_1^{1/3} T^{2/3} \quad (1)$$

and

$$R_{\rm mag} = 1.9 \times 10^6 \phi \mu_{26}^{4/7} m_1^{-1/7} \dot{M}_{17}^{-2/7} \quad (2)$$

where $m_1$ is the compact star mass in units of solar mass, $T$ is the compact star spin period in units of milliseconds, $\phi$ is the ratio between the magnetospheric radius and the Alfven radius, $\mu_{26}$ is the compact star magnetic moment in units of $10^{26}$ G cm$^3$ and $\dot{M}_{17}$ is the accretion rate in units of $10^{17}$ g s$^{-1}$. $R_{\rm co}$ and $R_{\rm mag}$ are given in cm. In this Letter, we assume that $\phi$ is almost independent of the accretion rate (Burderi & King 1998).

The requirements for X–ray pulsations (if there is no "intrinsic" pulse mechanism) and the presence of accretion flow (that is not centrifugally inhibited) give (see Li et al. 1999)

$$R_1 < R_{\rm mag}(\dot{M}_{\rm max}) < R_{\rm mag}(\dot{M}_{\rm min}) < R_{\rm co} \quad (3)$$

where $\dot{M}_{\rm min}$ and $\dot{M}_{\rm max}$ give the range of the accretion rate in which X–ray pulsations in SAX J1808.4–3658 were observed. From eqns. 1–3 we get (Li et al. 1999)

$$R_1 < 27.6 \left(\frac{F_{\rm max}}{F_{\rm min}}\right)^{-2/7} \left(\frac{T}{2.49 {\rm ms}}\right)^{2/3} m_1^{1/3} \text{ km} \quad (4)$$

where $F_{\rm max}$ and $F_{\rm min}$ are maximum and minimum values of measured X–ray fluxes respectively. It is to be noted that here the pulsar magnetic field is assumed to be dipolar. In writing eqn. 4, we also assume that $\dot{M}$ is proportional to the observed flux $F$ for all accretion rates. This is justified by the fact that the X–ray spectrum of SAX J1808.4–3658 was remarkably stable (Gilfanov et al. 1998), when the X–ray luminosity varied by a factor of $\sim 100$ during the 1998 April/May outburst. During this period, in 2−30 keV band, the maximum observed flux was around $3\times10^{-9}$ erg cm$^{-2}$ s$^{-1}$, while the flux value dropped to around $2\times10^{-11}$ erg cm$^{-2}$ s$^{-1}$ when the pulse signal became barely detectable (Cui, Morgan, & Titarchuk 1998). Adopting maximum to minimum flux ratio as 100 (after Li et al. 1999), we get from eqn. 4

$$R_1 < 7.40 m_1^{1/3} \text{ km} \tag{5}$$

---

[1] Joint Astronomy Program, Indian Institute of Science, Bangalore 560 012, INDIA; sudip@physics.iisc.ernet.in
[2] Indian Institute of Astrophysics, Bangalore 560 034, INDIA; sudip@iiap.ernet.in

Eqn. 5 gives the maximum value of $R_1$, if the maximum value of $m_1$ is known. To calculate $m_{1,\max}$ we first rewrite the eqn. 5 in the following form.

$$m_1 < 11.19 x_1^{-3/2} \tag{6}$$

where $x_1$ is the dimensionless radius to mass ratio of the compact star. We can compute $m_{1,\max}$ from eqn. 6, if the minimum value of $x_1$ is known. To choose the value of $x_{1,\min}$, we survey about 20 EOS (that include both SS and NS) and examine the value of $x_1$ corresponding to the maximum possible mass for a given EOS. For both SS and NS, we choose EOS of widely varying stiffness parameters, which guarantees our results to be of sufficient generality. This is reflected by the wide range of maximum possible mass values given in Table 1, where we have listed 13 representative EOS. From Table 1 and Figure 1, we notice that the $x_1$ values for all the EOS are confined to the range $2.98 - 4.34$, with 11 (out of 13) points clustering in $3.3 - 3.7$. To illustrate this, we draw a vertical line in Figure 1, corresponding to $x_1 = 2.9$. As none of the EOS points falls to the left of this line, we take 2.9 as the lower limit of $x_1$. Such a conclusion is very general, as it is valid for the whole range of existing EOS. This gives 2.27 (the crossing point of the vertical line and the curve in Figure 1) as the upper limit of $m_1$ from eqn. 6. The corresponding upper limit of $R_1$ comes out to be 9.73 km from eqn. 5.

It is to be noted that for some SS EOS, $x_1$ value may be less than 2.9 for lower values of masses (i.e., less than the maximum possible mass). But as we use the lower limit of $x_1$ to estimate the maximum possible value of $m_1$, it is justified to take 2.9 as the minimum possible value of $x_1$. An EOS model (that may be put forward in future) with $x_1$ (corresponding to the maximum possible mass) less than 2.9, will give a higher value of $m_{1,\max}$ than 2.27. However, such an unusual EOS is highly improbable. We also point out that if we take into account the rotation of the compact star, the lower limit of $x_1$ will increase, resulting in a decrease of $m_{1,\max}$. Therefore we can say that 2.27 may be the firm upper limit of $m_1$.

For the sake of completeness and to give more credibility to our work, we calculate $m_{1,\max}$ with less constraining values of $x_1$. For this purpose, we take $x_{1,\min} = 2.25$, which is the absolute lower limit (for compact star) of $x_1$ (Weinberg, S. 1972). This limit, which is independent of EOS and depends only on the structure of the relativistic equations for hydrostatic equilibrium, gives $m_{1,\max} = 3.32$ and $R_{1,\max} = 11.04$ km. Therefore 3.32 is the absolute upper limit of $m_1$. Another value of $x_{1,\min} (= 2.56)$ was derived by Bondi, H. (1964), under the reasonable assumptions concerning the EOS, i.e., $\epsilon > 0$, $p > 0$ and $dp/d\epsilon < 1$ (where $p$ is pressure and $\epsilon$ is energy density). This value of $x_{1,\min}$ implies $m_{1,\max} = 2.73$ and $R_{1,\max} = 10.34$ km. Therefore we see that the value of $R_{1,\max}$ is not very sensitive to the chosen value of $x_{1,\min}$.

## 3. LOWER MASS LIMIT

Here we estimate the probability ($P_{\min}$) of a possible compact star mass ($m_1$) to be the lower limit of mass. We do it using the procedure "Random distribution of orbital inclinations" for measuring neutron star mass, mentioned in Thorsett & Chakrabarty (1999). Because of the absence of sufficient observational data, here we can not follow any well-established statistical method. For example, the measured value of a single post–Keplerian (PK) parameter (Taylor, J.H. 1992) (with additional assumptions, such as a uniform prior likelihood for orbital orientations with respect to the observer) can be used to make strong statements about the posterior distribution of the masses (Thorsett & Chakrabarty 1999). But none of these parameters could be measured for SAX J1808.4–3658. Therefore, our results basically depend on the *a priori* probability of observing the source with a given inclination angle ($i$).

To explain the method, we first rewrite the well known expression for the pulsar mass function ($f_1$) in the following way.

$$\sin i = f_1^{1/3} \frac{(m_1 + m_2)^{2/3}}{m_2} \tag{7}$$

where $m_2$ is the mass of the companion star in units of solar mass. For a main sequence companion that fills its Roche lobe, $m_2 = 0.22$ (corresponding to $P_{\text{orb}} = 2.01$ hr). As a result, the lower limit of $i$ (i.e., $i_{\min}$) comes out to be $3^o$ from eqn. 7 (using $m_{1,\min} = 0$, the absolute lower limit). However Chakrabarty & Morgan (1998) have argued that $m_2 \lesssim 0.1$ (because the companion is bloated by irradiation). Therefore, we take 0.1 as the upper limit of $m_2$, that corresponds to $i_{\min} = 4^o$. The absence of a deep eclipse indicates that for a Roche lobe filling companion, $i_{\max} = 82^o$ (Chakrabarty & Morgan 1998). We set $m_{2,\min} = 0.05$, which is a possible companion mass according to Chakrabarty & Morgan (1998). We also take two other values of $m_{2,\min}$ for the purpose of illustration.

With all these limiting values, we calculate $P_{\min}$ in the following way. Given a value of $m_1$, we compute the allowed range of $i$ (i.e., $i_{a,\min}$ and $i_{a,\max}$) from eqn. 7, for the chosen range of $m_2$ (i.e., $m_{2,\min} \leq m_2 \leq m_{2,\max}$). However, for $i_{a,\max} > i_{\max}$, we take $i_{a,\max} = i_{\max}$. Similarly, we do not consider any value of $i$ less than $i_{\min}$. Now we argue (after Chakrabarty & Morgan 1998) that in statistical calculations it is useful to assume that binary orbits are randomly oriented with respect to the line of sight (see also Thorsett & Chakrabarty 1999). The differential distribution of inclinations is then proportional to $\sin i$. This gives the *a priori* probability of observing a system with $i$ in the range $i_{a,\min} \leq i \leq i_{a,\max}$ as $P = (\cos i_{a,\min} - \cos i_{a,\max})$. Therefore $P$ should be the probability of the chosen $m_1$ for being the actual compact star mass.

We calculate $P$ for many $m_1$ values (at regular interval) in the range $m_{1,\min} \leq m_1 \leq m_{1,\max}$. Then $P_{\min}$ is calculated by the formula

$$P_{\min} = \frac{\sum_{i=j}^{n} P_i}{\sum_{i=1}^{n} P_i} \tag{8}$$

where $P$ is calculated at $n$ number of $m_1$ points and $j$ is the index number of the $m_1$ value at which $P_{\min}$ is required.

In figure 2, we plot $P_{\min}$ against $m_1$ for three $m_{2,\min}$ values (0.04, 0.05 and 0.08). For $m_{2,\min} = 0.05$, we see that the minimum value of $m_1$ is 0.70 with 95% probability. Therefore, although we start with zero as the lower limit of $m_1$, we get a large value for $m_{1,\min}$ with a very high probability. This shows that the probabilistic method is very effective in estimating the value of $m_{1,\min}$. If we take $m_{1,\max} = 3.32$ (i.e., the absolute limit), the minimum value of $m_1$ comes out to be 0.90 with 95% probability, which shows that this method is sensitive to the assumed value of $m_{1,\max}$ (a less constrained value of the upper limit of $m_1$ gives a higher value of $m_{1,\min}$ with the same probability). A considerable increase in either $m_{2,\max}$ or $i_{\max}$ does not change our result much. For example, if $m_{2,\max} = 0.22$, the 95% probabilistic minimum value of $m_1$ is 0.68. However our result changes with the change of $m_{2,\min}$ for $m_{2,\min} < 0.04$.

We also point out that for a given value of $m_{2,\min}$, if $i_{\min}$ is greater than a certain value, it can be seen from eqn. 7 that every $i_{\min}$ will correspond to a minimum possible value of $m_1$. Therefore if we can observationally constrain $i$ from the lower side, the value of $m_{1,\min}$ can be predicted more accurately (as it will not depend on the probabilistic study). For example, the detailed modeling of the optical companion's multiband photometry during outburst with a simple X–ray heated disk model suggests that $\cos i < 0.45$ (Wang et al. 2001, in preparation; Bildsten & Chakrabarty 2001) for SAX J1808.4–3658. This implies $i > 63^o$ and hence $m_{1,\min} = 1.48$ (using $m_{2,\min} = 0.05$).

## 4. DISCUSSIONS

In this Letter, we have estimated the upper limits of the mass and the radius of the compact star in SAX J1808.4-3658. Li et al. (1999) have concluded that a narrow region in $m_1 - R_1$ space will be allowed for this star. The upper boundary of the mass will constrain this region effectively. It can also give the upper limit of $i$ (from eqn. 7), if $m_{2,\min}$ is known by an independent measurement. Alternatively, $m_{1,\max}$ gives the upper limit of $m_2$, for a known value of $i_{\min}$. For example, $i_{\min} = 63^o$ gives $m_{2,\max} = 0.066(0.085)$ for $m_{1,\max} = 2.27(3.32)$.

As we have mentioned in section 2, in this work, the pulsar magnetic field is assumed to be primarily dipolar. If the field has more complicated structure, the $R_{\rm mag} - \dot{M}$ relation will be changed, resulting in the modification of eqn. 4. This will lead to the change in eqn. 6, and hence our calculated value of $m_{1,\max}$ (and $R_{1,\max}$) will be modified. However, Li et al. (1999) have argued that the accretion flow around the compact star is dominated by a central dipole field, which gives credibility to our results.

Corresponding to every EOS, there exists a maximum possible mass. Therefore a lower limit of $m_1$ is very important in constraining EOS and hence in understanding the properties of matter at very high density compact star core. The possibility of this candidate to be a strange star can also be checked more effectively. Using our figure 2, we can predict the probability with which a certain value of $m_1$ will be the lower limit. For example, $m_1 = 1.41$ (the maximum possible mass for our model Y) will be the lowest possible mass with 72% probability (from curve 1 of Figure 2). However, it is to be kept in mind that such a probabilistic study may be valid, if binary inclination angles are distributed randomly.

If the orbital evolution of SAX J1808.4-3658 is driven by only gravitational wave radiation, then the rate of change of the orbital period will be given by (Ergma & Antipova 1999)

$$\dot{P}_{\rm orb} = -1.72 \times 10^{-7}(2\pi/P_{\rm orb})^{5/3}m_2 m_1(m_1+m_2)^{-1/3} \quad (9)$$

This implies $2.30 \times 10^{-13}$ as the maximum possible (absolute) value of $\dot{P}_{\rm orb}$ (for $m_{1,\max} = 2.27$, $m_{2,\max} = 0.1$ and $P_{\rm orb} = 7249$ s). Chakrabarty & Morgan (1998) have suggested that $\dot{P}_{\rm orb}$ can be measured, if the source remains in the X–ray bright state for long enough (or if the pulsations remain detectable in quiescence). If in future, such a measurement yields the value of the orbital period decay rate greater than $2.30 \times 10^{-13}$ ($2.99 \times 10^{-13}$ for $m_{1,\max} = 3.32$), then we can conclude with certain confidence that the orbital evolution of SAX J1808.4-3658 is significantly driven by magnetic braking. This will give support to the evolutionary model of Ergma & Antipova (1999) and in general will be very important for learning about the prehistory of the system. A better understanding of the criterion for magnetic braking will also be possible.

It has been proposed that SAX J1808.4-3658 may emerge as a radio pulsar during the X–ray quiescence (Chakrabarty & Morgan 1998). Ergma & Antipova (1999) have calculated that for $\lambda < 3$ cm, it may be possible to observe radio emission from this source. However our limits of mass values give a slightly higher (3.8 cm for $m_{1,\max} = 2.27$ and 4.5 cm for $m_{1,\max} = 3.32$) upper limit for $\lambda$.

As we have already mentioned in section 3, a moderately high value of $i_{\min}$ will give a lower limit of $m_1$ (without any probabilistic study). This will be very important for constraining EOS more decisively. For example, if $i_{\min} = 63^o$ (corresponds to $m_{1,\min} = 1.48$, given in section 3), our EOS models SS1, SS2 and Y will be unfavoured (see $m_{1,\max}$–column of Table 1). According to Chakrabarty & Morgan (1998), a deeper eclipse might be observed for the less penetrating radio emission, providing a strong constraint on the value of $i$. Therefore, we expect that, the value of $i_{\min}$ (determined by this method) may be able to rule out several soft EOS models in future.

We acknowledge Dipankar Bhattacharya for reading the manuscript and giving valuable suggestions. We thank Xiangdong Li for sending useful comments and Pijush Bhattacharjee for encouragement. We also thank the referee for giving suggestions to improve the quality of the paper.

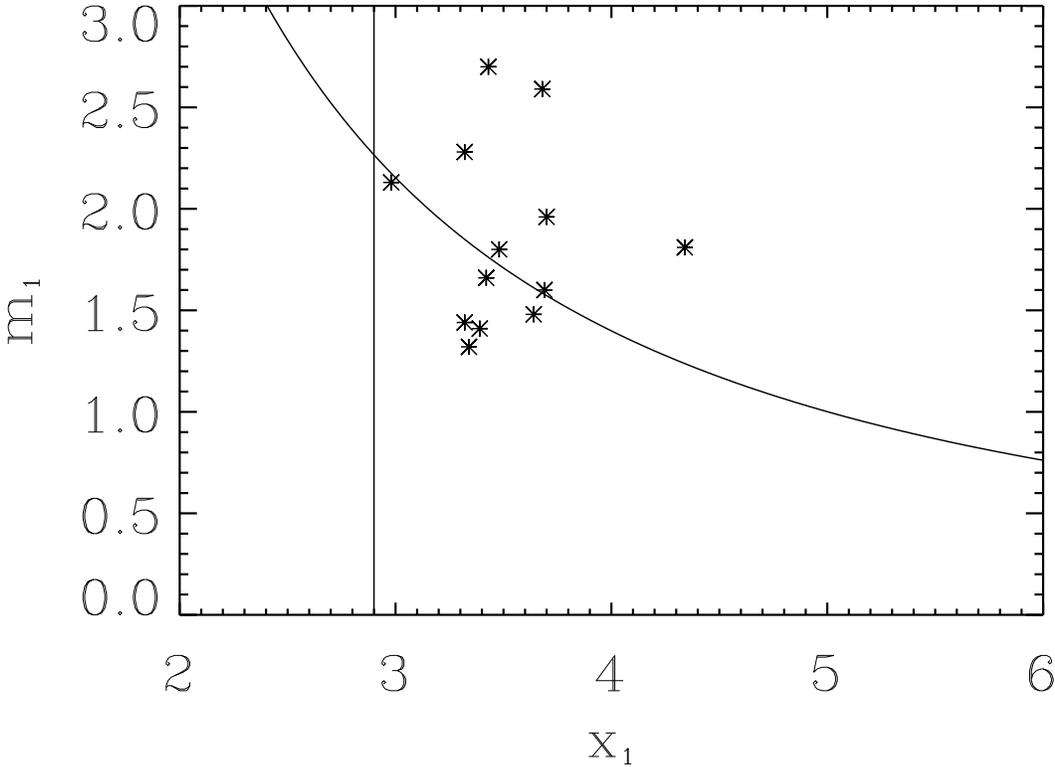

FIG. 1.— $m_1$ vs. $x_1$ plot (see the text) for compact star. The solid curve indicates the upper bound of the mass according to eqn. 6. The asterisks are for different EOS models listed in Table 1. The vertical line corresponds to $x_1 = 2.9$.

TABLE 1
List of 13 EOS (both SS and NS) of widely varying stiffness parameters and their references. The values of relevant properties (see the text) are also given.

| EOS label | compact star | Reference | $m_{1,\max}$ | $R_1(km)$ | $x_1$ |
|---|---|---|---|---|---|
| SS2 | SS | Dey et al. (1998) | 1.32 | 6.53 | 3.34 |
| SS1 | SS | Dey et al. (1998) | 1.44 | 7.07 | 3.32 |
| $B_{90}$ | SS | Farhi & Jaffe (1984), $B = 90$ MeV/fm$^3$, $m_s = 0$ | 1.60 | 8.74 | 3.69 |
| $B_{60}$ | SS | Farhi & Jaffe (1984), $B = 60$ MeV/fm$^3$, $m_s = 0$ | 1.96 | 10.71 | 3.70 |
| Y | NS | Pandharipande (1971b), hyperonic matter | 1.41 | 7.10 | 3.39 |
| B | NS | Baldo, Bombaci, & Burgio (1997), nuclear matter | 1.79 | 9.64 | 3.64 |
| W | NS | Walecka (1974), neutron matter | 2.28 | 11.22 | 3.32 |
| SBD | NS | Sahu, Basu, & Datta (1993), nuclear matter | 2.59 | 14.08 | 3.68 |
| A | NS | Pandharipande (1971a), Reid soft core | 1.66 | 8.37 | 3.42 |
| AU | NS | Wiringa, Fiks, & Fabrocini (1988), AV14 + UVII | 2.13 | 9.41 | 2.98 |
| FPS | NS | Lorenz, Ravenhall, & Pethick (1993), UV14 + TNI | 1.80 | 9.28 | 3.48 |
| L | NS | Pandharipande & Smith (1975b), mean field | 2.70 | 13.70 | 3.43 |
| M | NS | Pandharipande & Smith (1975a), tensor interaction | 1.81 | 11.60 | 4.34 |

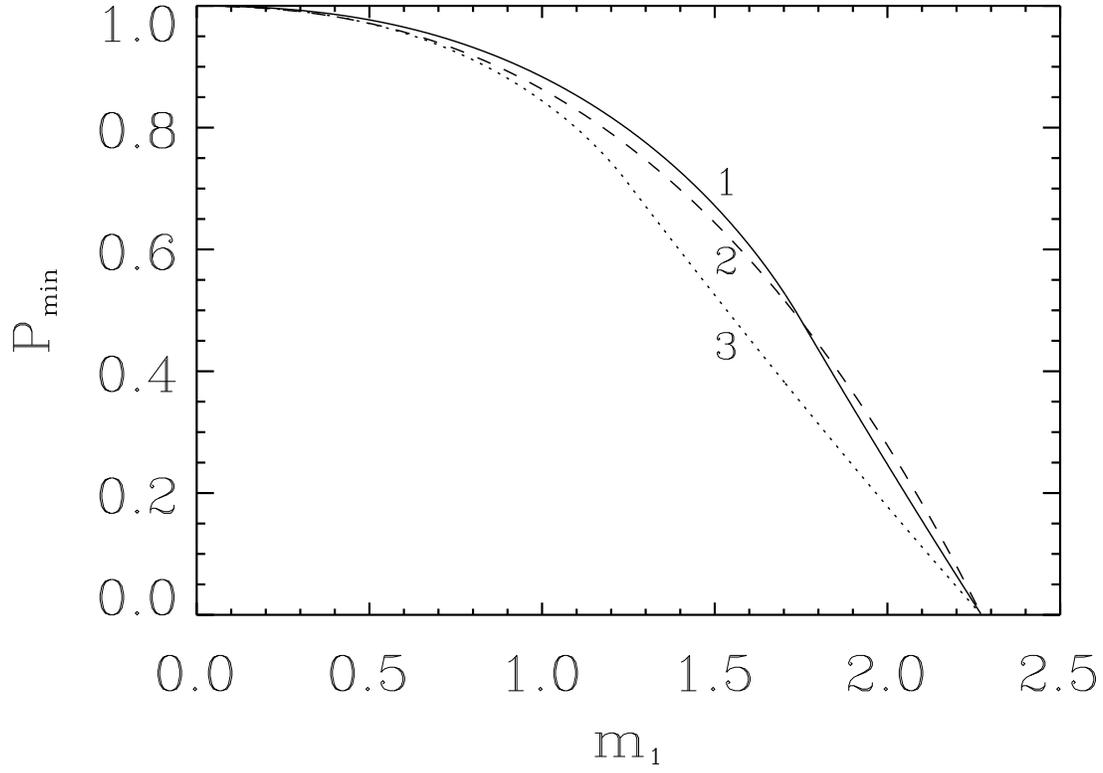

Fig. 2.— $P_{\min}$ vs. $m_1$ plot with the following parameter values : $m_{1,\min} = 0$, $m_{1,\max} = 2.27$, $m_{2,\max} = 0.1$, $i_{\min} = 4^o$ and $i_{\max} = 82^o$. Curve 1 is for $m_{2,\min} = 0.05$, curve 2 for $m_{2,\min} = 0.08$ and curve 3 for $m_{2,\min} = 0.04$.